\documentclass[12pt,a4paper]{article}
\usepackage[english]{babel}
\usepackage{amsmath,natbib,bookman}
\usepackage[dvips]{graphicx}
\usepackage[footnotesize,bf]{caption2}
\usepackage{textcomp}
\usepackage{amsfonts,lscape,epsfig}
\title{Unexpected volatility and intraday serial 
correlation\footnote{We acknowledge participants at the {\it IV Workshop LABSI},
  Siena, and Taro Kanatani for useful comments. 
SB thankfully acknowledges the Welch foundation for financial
  support through Grant no. B-1577.}}
\author{Simone Bianco \\ {\it Center for Nonlinear Science, University
  \it of North Texas}\\ {\it P.O. Box 311427, Denton, Texas, 76201-1427}
  \\ {\it e-mail: {\tt sbianco@unt.edu}}\\ \vspace{0.7cm} \\ Roberto Ren\`o
\\{\it Dipartimento di Economia Politica, Universit\`a di Siena} \\
{\it Piazza S.Francesco 7, 53100, Siena} \\ {\it e-mail: {\tt reno@unisi.it}} }

\begin{document}
\maketitle

\begin{abstract}
We study the impact of volatility on intraday serial correlation, at
time scales of less than $20$ minutes, 
exploiting a data set with all transaction on
SPX500 futures from 1993 to 2001. We show that, while realized volatility
and intraday serial correlation are linked, this relation is driven by
unexpected volatility only, that is by the fraction of volatility which
cannot be forecasted. The impact of predictable volatility is instead found to
be negative (LeBaron effect). Our results are robust to microstructure noise,
and they confirm the leading economic theories on price formation.
\end{abstract}

\newpage
\section{Introduction}
The study of serial correlation in asset prices 
is of great importance in financial
economics. Indeed, from the point of view of market efficiency
\citep{Fam70}, as well as market inefficiency \citep{Shl03},
serial correlation is a market anomaly which need to be addressed by 
economic theories.
Once serial correlation is significantly detected in the data, see
\citet{Jam03} as an example, an
explanation is needed to reconcile the empirical finding with the
assumption of informational efficiency of the market. 
This has been typically accomplished in
a rational setting \citep{LoMac90,BouRicWhi94,SenWad92,Saf00} or in a
behavioral setting
\citep{CutPotSum91,JegTit93,Cha93,BadKalNoe95,ChaGal05}.  
In this paper, we concentrate on very short-run serial correlation,
that is we focus on intraday data and in particular on time scales
from $4$ to $20$ minutes. 

The purpose of this paper is
multiple. Beyond showing the informational efficiency of the
considered market, which is actually out of discussion given its
liquidity, 
our aim is to study the dynamical properties of
intraday serial correlation. We extend previous literature by
decomposing intraday volatility, measured by means of realized
volatility, into its predictable and unpredictable part.
To quantify intraday serial correlation, we use the variance-ratio test
on evenly sampled intraday data. While being very standard for daily
data, the variance ratio test has still little application on
high-frequency data, including \citet{AndBolDas01,ThoPat03,KauSap05}. 

Our main result is that intraday serial
correlation is positively
linked with {\it unexpected volatility}, defined as the
residual in a linear regression model for daily volatility as measured
with intraday data. In other words, unexpected volatility is that part
of volatility which was not forecasted on that market in that
particular day.
We also explain the puzzling results of \citet{BiaRen06} who, on a much
less liquid market (Italian stock index futures), found
volatility to be positively correlated with serial correlation, at
odds with the result in \citet{Leb92}. We show that indeed total
volatility is positively related to serial correlation: however, 
it is unexpected volatility that drives this positive
relation. The predictable part of volatility, that  used in
\citet{Leb92}, turns out to be negatively related to serial
correlation, in agreement with previous literature.

The paper is organized as follows. Section \ref{data} illustrates the
methodology and describes the data set. Section \ref{results} shows
the estimation results and discusses the implications of
them. Section \ref{conclusions} concludes.
 
\section{Data and methodology}\label{data}

The data set under study is the collection of all transactions on the
S\&P500 stock index futures from April, 1993 to October 2001, for a
total of 1,975 trading days. We have information on all futures
maturity, but we use only next-to-expiration contracts, with the S\&P
500 expiring quarterly. We use only transactions from $8:30$ a.m. to $3:15$
p.m..
In total, we have $4,898,381$ transactions,
that is $2,480$ per day on average, with an average duration between
adjacent trades of $9,8$ seconds. 
Not all high-frequency information is used. We use instead a grid of
evenly sampled data every day. We find that a time interval of $\Delta
t=4$ minutes is a large enough to avoid the
problem of intervals with no price changes within. Thus, for every
day, we have a time series of $101$ evenly sampled prices.

To study intraday serial correlation, we use the variance-ratio
statistics. This briefly consists in what follows.
Denote by
$P_k,~k=1,\ldots, N$ a time series and 
define the first differences time series
$r_k=P_k-P_{k-1}$. The variance ratio at lag $q$ is given by 
 \begin{equation}\label{vrq}
  VR(q)=\frac{Var[r_k(q)]}{Var[r_k]} 
\end{equation}
 where
\begin{equation}
  r_k(q) = \sum_{j=1}^{q+1} r_{k+j} 
\end{equation}
represents the $q-$period return.
We implement the variance ratio test according to the
heteroskedastic consistent estimator \citep{LoMac88} with overlapping
observations \citep{RicSmi93}, for which the asymptotic distribution
is well known under the null, see Appendix \ref{appendix}.
In particular, \citet{BiaRen06} show that the VR test can be
implemented on high frequency data of stock index futures 
transactions, for time scales lower than $20$
minutes, given the typical heteroskedasticity of this asset. 
This is in line with the robustness analysis of \citet{DeoRic03}.
We then study values of $q$ ranging from $1$ to $5$, since in our case
the interval between adjacent observations is $4$ minutes.
For these values of $q$, we can then safely use the VR test with
high-frequency data in our context. 

We then compute $1,975$ daily values of the variance ratio for
$q=1,\ldots,5$. The top panel of table \ref{violations} reports the number of
significantly positive and negative variance ratios, for different
confidence intervals. The positive violations are compatible with the null. The excess in negative
violations can instead be ascribed to the bid-ask bounce effect, see
the thorough discussion in \citet{BiaRen06}.

\begin{figure}[p]
\begin{center}
\mbox{\epsfig{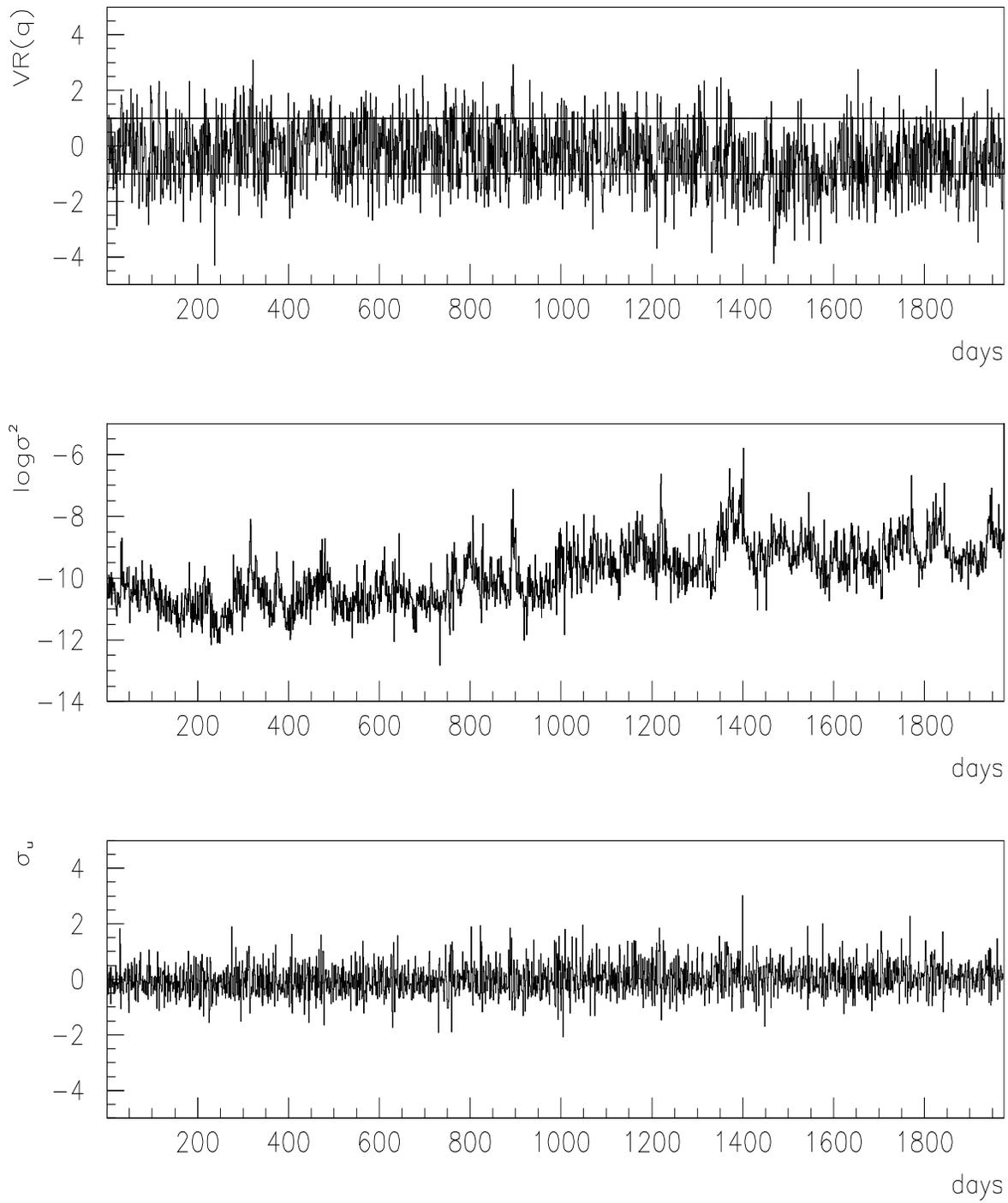}}
\caption{From top to bottom: the time series of
  $\widetilde{VR}(1)$ with one standard deviation bands,
  the daily realized volatility and the estimated unexpected
  volatility.}\label{vrts}
\end{center}
\end{figure}

In order to quantify the daily serial correlation, we use the
standardized variance ratio at different lags $q$, defined as: 
\begin{equation}\label{stdvr}
\widetilde{VR}(q) = \sqrt{nq}\frac{\widehat{VR}(q) -
  1}{\sqrt{\hat{\theta}(q)}},  
\end{equation}
where $\hat{\theta}(q)$ is the heteroskedastic consistent estimator of
the variance ratio variance, see Appendix \ref{appendix}.
The time series of variance ratios at $q=1$ is shown in figure \ref{vrts} 

Given the high persistence in volatility, also the standardized
variance ratio is found to be highly persistent. We discuss further
this point in Section \ref{results}.

We want to link serial correlation with volatility. On each day, in
which we have $N$ returns, we
define volatility as 
\begin{equation}\label{dvdt}
\sigma^2 = \displaystyle \sum_{k=1}^{N} r_k^2
\end{equation}
This is the well-known measure of realized variance,
see \citet{AndBolDie03}.
However, in what follows 
we argue that an other variable plays a very special role,
that is unexpected volatility. We know that volatility is highly
foreseeable in financial markets, see \citet{PooGra03} for a review,
mainly given its persistence. Moreover, a simple linear model for
realized volatility leads to fair forecasts, see e.g. \citet{ABDL03,Coretal01}. 
We then assume that the market
volatility is forecasted with the following linear model:
\begin{equation}\label{dsigma}
  \log(\sigma^2_t) = \alpha + \beta_1 \log(\sigma^2_{t-1}) + \beta_2
  \log(\sigma^2_{t-2}) + \beta_3 \log(\sigma^2_{t-3}) +  \varepsilon_t.
\end{equation}

\noindent Even if the model (\ref{dsigma}) is fairly simple, since it ignores
long-memory and leverage effects, on the US stock
index futures data it yields an $R^2$ of $66.2\%$. 
We then define {\it unexpected volatility} as the residuals of the
above regression, 
\begin{equation}
\sigma_{u,t} \equiv
\hat{\varepsilon}_t.\end{equation} 
We also define the predictable part of volatility, as:
$$\sigma_{p,t} \equiv \log(\sigma^2_t)-\sigma_{u,t}$$
By construction, lagged volatility at times
$t-1,t-2,t-3$ and unexpected volatility are orthogonal. Thus
$\sigma_{p,t}$ and $\sigma_{u,t}$ are orthogonal as well.

It is clear that our definition of unexpected and predictable volatility
is dependent on model (\ref{dsigma}); however the inclusion of further
lags does not change our results; and including more complicated
effects does not improve the specification of model
\ref{dsigma}, see the extensive study of \citet{HanLun05}.
Also nonlinear specifications, as those of
\citet{MahMcc02}, have been found to 
yield forecast improvements which are not substantial.

\section{Results}\label{results}
We start from the finding in \citet{BiaRen06} that standardized variance
ratios are negatively autocorrelated, and we confirm this finding on US data.
However, this feature is inherited by the serial
auto-correlation of the volatility itself. To check this, we simulate
a long series of a GARCH(1,1) process with zero auto-correlation. On
the simulated series we spuriously detect an autocorrelated
standardized variance ratio. Since the simulated series is
persistent, we conclude that 
the serial correlations of the standardized VRs is a consequence
of the heteroskedasticity of the data. However, in order to get reliable
specification when the variance ratio is the dependent variable, it is
necessary to add lagged variance ratio regressors as explanatory variables.

As an overall specification test for the regression,
we use the Ljung-Box test of residuals at lag $5$ and we denote it by
$Q(5)$.
We first study a model in which we include volatility as a
regressor:
\begin{equation}\label{model2}
  \widetilde{VR}_t = \alpha 
+ \sum_{i=1}^4 \delta_i
  \widetilde{VR}_{t-i} 
+ \beta \cdot \log(\sigma^2_t) 
+ \varepsilon_t. 
\end{equation}
Results are in Table \ref{table-ar-sigma}. 
We find that there is a positive and significant relation between
volatility and standardized variance ratio, and the regression is well
specified if we include enough autoregressive terms for the variance ratio,
see the Ljung-Box statistics. 
This result is not entirely surprising. On a much smaller market
(Italy), \citet{BiaRen06} provide evidence of a positive relation
between volatility and intraday serial correlation. This is different
from what is 
typically found at daily level, where the correlation is found
to be negative, according to the LeBaron effect
\citep{Leb92,SenWad92}. However, this result can be explained
according to the model of reinforcement of opinions of
\citet{Cha93}. According to this model, serial correlation is
introduced into data since once an investor decides to buy, he
observes more liquid substitutes and reinforce his opinion according to the
movements of the substitutes. This effect is stronger when volatility
is high, that is when the price move more (or more rapidly). Thus, the
\citet{Cha93} model posits a positive relation between volatility and
intraday serial correlation which is at all reasonable. However, for
the US market the Chan model is less tenable. Indeed,
for the US it is unreasonable to look for a
more liquid substitute. Thus, the effect of the reinforcement of
opinions is likely to be milder.
To better understand this, we compute the percentage of significant
VRs as volatility increases. The violations are reported in Table
\ref{violations}. On the contrary on what happens on the Italian market,
where the percentage of positive violations increases when volatility
increases, we find that this holds marginally for the US market,
confirming our intuition that the mechanism of reinforcement of
opinions is likely to play 
a minor role in a liquid market as the US stock index
futures.

We then analyze the impact of unexpected volatility.
We estimate the regression:
\begin{equation}\label{model3}
\widetilde{VR}_t = \alpha + \sum_{i=1}^4 \delta_i
  \widetilde{VR}_{t-i} + \beta \cdot \sigma_{u,t} + \varepsilon_t. 
\end{equation}
Results are shown in Table \ref{table-ar-sigmau}. Unexpected volatility is
found to be highly significant, and we obtain a good specification as measured by the Ljung-Box
statistics, as far as we include enough lags of the variance ratio
itself and $q$ is large enough.
Thus, it is evident that unexpected volatility plays a crucial role in the
emergence of intraday serial correlations, for all the considered time
scales. 

Most importantly, our results can be reconciled with the results in
\citet{Leb92}. To show this, we estimate 
the encompassing regression:
\begin{equation}\label{model4}
  \widetilde{VR}_t = \alpha + \sum_{i=1}^4 \delta_i\widetilde{VR}_{t-i} + \beta \cdot \sigma_{p,t} + \gamma \cdot \sigma_{u,t} + \varepsilon_t,
\end{equation}
where both unexpected and predictable volatility are included as regressors.
Results are displayed in Table \ref{table-ar-sigma-sigmau} and
indicate that, while volatility has been found to be significant in
model (\ref{model2}), its predictable part is negatively related with
intraday variance ratios, and its unexpected part is positively
related. Indeed, \citet{Leb92} did not use
realized measures of intraday variance, but he filtered the variance
with a GARCH-like model, thus he considered only the predictable
part, getting a negative relation. Since we are using a realized
measure of volatility, we can decompose it into a predictable and
unpredictable part, and we consistently find that the first has a
negative impact on intraday serial correlation, while the second has a
large positive impact. A negative relation between predictable
volatility and intraday serial correlation could not be seen by
\citet{BiaRen06} in the Italian market, given the very low statistics
(three years of data only).
Thus, we conclude that unexpected
volatility is the main source of intraday serial correlation, even if,
at our knowledge, there is not an economic model explaining why the
role of unexpected volatility is so important, since most economic
models use total volatility.

\section{Conclusions}\label{conclusions}
In this paper
we study the impact of volatility on
intraday serial correlation in the US stock index futures market, which is the
most liquid market in the world.     
We exploit the availability of intraday data to measure volatility by
means of realized variance, and intraday serial correlation by means
of standardized variance ratio.
We find that, in agreement with the economic theory, total volatility
plays a minor role in the US market, since the mechanism of
reinforcement of opinions postulated by \citet{Cha93} is less important
in this market. We then use our realized measure to decompose
volatility into its predictable and unpredictable part, which we call
unexpected volatility.
We extend previous findings in the literature in the following direction.
We find that there is a positive and significant relation between
unexpected volatility and intraday serial correlation, while we
confirm the LeBaron effect: predictable volatility is negatively
related to serial correlation.

This result can be important for the economic theory, since this could
potentially reveal basic properties about the pricing formation
mechanism. As far as we know, there are no economic theories
explaining the stylized fact documented by our study, thus our results 
introduce a new challenge. 
However, we presume that the role of unexpected volatility
is linked to the way information is spread in the market. In this
respect, unexpected volatility could be potentially employed as a proxy for 
information asymmetry. Further research is needed to assess this
conjecture.  

\bibliographystyle{Chicago}
\bibliography{/home/reno/specials/Reno.bib}

\begin{table}[!tb]  
  \begin{center} 
\vskip 1pt 

all $\sigma^2, 100\%$ of the sample \\ \vskip 1pt
    \begin{tabular}{p{0.4cm}p{1.4cm}p{1.4cm}p{1.4cm}p{1.4cm}p{1.4cm}p{1.4cm}}  
      \hline   
      $q$ & $90\% ^+$ & $90\% ^-$ & $95\% ^+$ & $95\% ^-$ & $99\% ^+$ & $99\% ^-$ \\   
      \hline
       1 &  0.064 &  0.209 &  0.034 &  0.120 &  0.004 & 0.034 \\
       2 &  0.057 &  0.202 &  0.029 &  0.103 &  0.005 & 0.021 \\
       3 &  0.051 &  0.191 &  0.025 &  0.093 &  0.006 & 0.013 \\
       4 &  0.047 &  0.173 &  0.024 &  0.065 &  0.006 & 0.003 \\
       5 &  0.044 &  0.157 &  0.024 &  0.053 &  0.006 & 0.001 \\
       \hline \end{tabular}
\vskip 1pt 
$\sigma^2 > 3\cdot 10^{-5}$, 68.5 \% of the sample \\ \vskip 1pt
    \begin{tabular}{p{0.4cm}p{1.4cm}p{1.4cm}p{1.4cm}p{1.4cm}p{1.4cm}p{1.4cm}} 
\hline 
      $q$ & $90\% ^+$ & $90\% ^-$ & $95\% ^+$ & $95\% ^-$ & $99\% ^+$ & $99\% ^-$ \\   \hline
1 &   0.089 &   0.205 &   0.045 &   0.123 &   0.004 &   0.039\\ 
2 &   0.069 &   0.201 &   0.031 &   0.109 &   0.007 &   0.024\\ 
3 &   0.066 &   0.177 &   0.030 &   0.098 &   0.006 &   0.013\\ 
4 &   0.055 &   0.161 &   0.030 &   0.072 &   0.007 &   0.001\\ 
5 &   0.052 &   0.153 &   0.027 &   0.061 &   0.007 &   0.001\\ 
\hline\end{tabular}
\vskip 1pt 
$\sigma^2 > 7.5\cdot 10^{-5}$, 35.8 \% of the sample \\ \vskip 1pt
    \begin{tabular}{p{0.4cm}p{1.4cm}p{1.4cm}p{1.4cm}p{1.4cm}p{1.4cm}p{1.4cm}}  
      \hline   $q$ & $90\% ^+$ & $90\% ^-$ & $95\% ^+$ & $95\% ^-$ &
      $99\% ^+$ & $99\% ^-$ \\   \hline
1 &   0.088 &   0.212 &   0.041 &   0.127 &   0.007 &   0.048\\ 
2 &   0.059 &   0.213 &   0.027 &   0.113 &   0.007 &   0.021\\ 
3 &   0.055 &   0.185 &   0.021 &   0.100 &   0.007 &   0.013\\ 
4 &   0.042 &   0.154 &   0.025 &   0.068 &   0.007 &   0.001\\ 
5 &   0.042 &   0.145 &   0.025 &   0.058 &   0.008 &   0.000\\ 
\hline\end{tabular}
%
\vskip 1pt 
$\sigma^2 > 1.4\cdot 10^{-4}$, 15.8 \% of the sample \\ \vskip 1pt
    \begin{tabular}{p{0.4cm}p{1.4cm}p{1.4cm}p{1.4cm}p{1.4cm}p{1.4cm}p{1.4cm}}  

      \hline   $q$ & $90\% ^+$ & $90\% ^-$ & $95\% ^+$ & $95\% ^-$ &
      $99\% ^+$ & $99\% ^-$ \\   \hline
1 &   0.128 &   0.147 &   0.071 &   0.096 &   0.016 &   0.035\\ 
2 &   0.093 &   0.183 &   0.045 &   0.096 &   0.013 &   0.022\\ 
3 &   0.074 &   0.173 &   0.038 &   0.093 &   0.010 &   0.013\\ 
4 &   0.061 &   0.138 &   0.035 &   0.067 &   0.010 &   0.003\\ 
5 &   0.058 &   0.154 &   0.035 &   0.064 &   0.010 &   0.000\\ 
\hline\end{tabular}
\vskip 1pt 
$\sigma^2 > 2\cdot 10^{-4}$, 8.1 \% of the sample \\ \vskip 1pt
    \begin{tabular}{p{0.4cm}p{1.4cm}p{1.4cm}p{1.4cm}p{1.4cm}p{1.4cm}p{1.4cm}}  

      \hline   $q$ & $90\% ^+$ & $90\% ^-$ & $95\% ^+$ & $95\% ^-$ &
      $99\% ^+$ & $99\% ^-$ \\   \hline
1 &   0.151 &   0.132 &   0.094 &   0.082 &   0.025 &   0.038\\
2 &   0.119 &   0.164 &   0.069 &   0.101 &   0.013 &   0.019\\
3 &   0.101 &   0.151 &   0.057 &   0.094 &   0.013 &   0.013\\
4 &   0.069 &   0.132 &   0.038 &   0.069 &   0.013 &   0.006\\
5 &   0.063 &   0.138 &   0.038 &   0.069 &   0.013 &   0.000\\

\hline\end{tabular}
\caption{Percentage of significant positive and negative VR, for
  different significance levels (one-sided), 
on subsamples with growing daily volatility, see the top of each panel.}\label{violations}
\end{center}
\end{table}

\begin{table}[!tb]
\tiny
  \begin{center}
    \begin{tabular}{|c|c|c|c|c|c|c|c|}
      \hline
      $q$ & $\alpha$ &$\log(\sigma^2)$  & $\widetilde{VR}_{t-1}$ &
      $\widetilde{VR}_{t-2}$ & $\widetilde{VR}_{t-3}$ & $\widetilde{VR}_{t-4}$
      & Q(5)  \\
      \hline
      1 &    0.512   &    0.088  &        &    &    &    &     201.88$^{**}$ \\
      &  (  1.982  )$^{*}$ &   (  3.389  )$^{**}$ &        &    &    &    &       \\\cline{2-8}
      &    0.610   &    0.092   &     0.179   &    &    &    &      67.66$^{**}$ \\
      &  (  2.398  )$^{**}$ &   (  3.575  )$^{**}$ &   (  8.100  )$^{**}$ &    &    &    &       \\\cline{2-8}
      &    0.741   &     0.102  &     0.161   &     0.100   &    &    &      37.73$^{**}$ \\
      &   (  2.905  )$^{**}$ &   (  3.975  )$^{**}$ &   (  7.200  )$^{**}$ &   (  4.464  )$^{**}$ &    &    &       \\\cline{2-8}
      &    0.819   &     0.108  &     0.153   &     0.086   &     0.088   &    &      18.30$^{**}$ \\
      &   (  3.210  )$^{**}$ &   (  4.196  )$^{**}$ &   (  6.824  )$^{**}$ &   (  3.808  )$^{**}$ &   (  3.912  )$^{**}$ &    &       \\\cline{2-8}
      &    0.881   &     0.112   &    0.148   &     0.082   &     0.078   &     0.068  &       6.89 \\
      &   (  3.452  )$^{**}$ &   (  4.380  )$^{**}$ &   (  6.590  )$^{**}$ &
      (  3.611  )$^{**}$ &   (  3.462  )$^{**}$ &   (  3.014  )$^{**}$ &
      \\ 
      \hline
      2 &    0.070   &     0.049   &        &    &    &    &     120.56$^{**}$ \\
      &   (  0.295  ) &   (  2.022  )$^{*}$ &    &    &    &    &       \\\cline{2-8}
      &    0.193   &     0.055   &     0.147  &    &    &    &      43.73$^{**}$ \\
      &   (  0.817  ) &   (  2.309  )$^{*}$ &   (  6.585  )$^{**}$ &    &    &    &       \\\cline{2-8}
      &    0.306   &     0.064   &     0.136   &     0.075   &    &    &      29.09$^{**}$ \\
      &   (  1.284  ) &   (  2.665  )$^{**}$ &   (  6.065  )$^{**}$ &   (  3.315  )$^{**}$ &    &    &       \\\cline{2-8}
      &    0.445   &     0.074   &     0.129   &     0.062   &     0.103   &    &       4.67 \\
      &   (  1.859  )$^{*}$ &   (  3.108  )$^{**}$ &   (  5.779  )$^{**}$ &   (  2.741  )$^{**}$ &   (  4.590  )$^{**}$ &    &       \\\cline{2-8}
      &    0.498   &     0.078   &     0.125   &     0.060   &     0.098   &    0.043   &       0.97 \\
      &   (  2.068  )$^{*}$ &   (  3.271  )$^{**}$ &   (  5.578  )$^{**}$ &   (  2.671  )$^{**}$ &   (  4.338  )$^{**}$ &   (  1.901  )$^{*}$ &       \\
      \hline
      3 &    0.073   &     0.050   &        &    &    &    &      62.86$^{**}$ \\
      &   (  0.321  ) &   (  2.216  )$^{*}$ &    &    &    &    &       \\\cline{2-8}
      &    0.144   &     0.053   &     0.102   &    &    &    &      27.50$^{**}$ \\
      &   (  0.641  ) &   (  2.353  )$^{**}$ &   (  4.559  )$^{**}$ &    &    &    &       \\\cline{2-8}
      &    0.210   &     0.058   &     0.097   &     0.050   &    &    &      20.37$^{**}$ \\
      &   (  0.923  ) &   (  2.547  )$^{**}$ &   (  4.317  )$^{**}$ &   (  2.235  )$^{*}$ &    &    &       \\\cline{2-8}
      &    0.312   &     0.065   &     0.093   &     0.043   &     0.083   &    &       3.75 \\
      &   (  1.370  ) &   (  2.865  )$^{**}$ &   (  4.162  )$^{**}$ &   (  1.911  )$^{*}$ &   (  3.676  )$^{**}$ &    &       \\\cline{2-8}
      &    0.351   &     0.068   &     0.091   &     0.043   &     0.080   &     0.032   &       1.15 \\
      &   (  1.529  ) &   (  2.981  )$^{**}$ &   (  4.054  )$^{**}$ &   (
      1.898  )$^{*}$ &   (  3.551  )$^{**}$ &   (  1.401  ) &       \\
      \hline
      4 &    0.199   &     0.063   &        &    &    &    &      26.03$^{**}$ \\
      &   (  0.917  ) &   (  2.881  )$^{**}$ &    &    &    &    &       \\\cline{2-8}
      &    0.225   &     0.063   &     0.054   &    &    &    &      15.13$^{**}$ \\
      &   (  1.040  ) &   (  2.902  )$^{**}$ &   (  2.391  )$^{**}$ &    &    &    &       \\\cline{2-8}
      &    0.251   &     0.065   &     0.052   &     0.027   &    &    &      12.59$^{*}$ \\
      &   (  1.150  ) &   (  2.961  )$^{**}$ &   (  2.323  )$^{*}$ &   (  1.192  ) &    &    &       \\\cline{2-8}
      &    0.311   &     0.069   &     0.051   &     0.024   &     0.056   &    &       3.88 \\
      &   (  1.420  ) &   (  3.134  )$^{**}$ &   (  2.266  )$^{*}$ &   (  1.076  ) &   (  2.469  )$^{**}$ &    &       \\\cline{2-8}
      &    0.340   &     0.071   &     0.050   &     0.025   &     0.055   &     0.025   &       1.62 \\
      &   (  1.541  ) &   (  3.215  )$^{**}$ &   (  2.215  )$^{*}$ &   (
      1.104  ) &   (  2.419  )$^{**}$ &   (  1.117  ) &       \\
      \hline
      5 &    0.269   &     0.069   &        &    &    &    &       8.24 \\
      &   (  1.277  ) &   (  3.256  )$^{**}$ &    &    &    &    &       \\\cline{2-8}
      &    0.272   &     0.069   &     0.007   &    &    &    &       7.65 \\
      &   (  1.287  ) &   (  3.251  )$^{**}$ &   (  0.317  ) &    &    &    &       \\\cline{2-8}
      &    0.272   &     0.069   &     0.007   &     0.005   &    &    &       7.55 \\
      &   (  1.284  ) &   (  3.237  )$^{**}$ &   (  0.313 )  &   (  0.201  ) &    &    &       \\\cline{2-8}
      &    0.308   &     0.071   &     0.007   &     0.004   &     0.036   &    &       3.26 \\
      &   (  1.446  ) &   (  3.332  )$^{**}$ &   (  0.315  ) &   (  0.194  ) &   (  1.586  ) &    &       \\\cline{2-8}
      &    0.329   &     0.073   &     0.006   &     0.006   &     0.036   &     0.020   &       1.55 \\
      &   (  1.537  ) &   (  3.390  )$^{**}$ &   (  0.284  ) &   (  0.252  ) &   (  1.588  ) &   (  0.892  ) &       \\
      \hline
    \end{tabular}
\caption{Estimates of model \ref{model2}), for different values of $q$. $^{*}$ indicates $95\%$ of confidence level, $^{**}$ $99\%$ of
      confidence level.}\label{table-ar-sigma}
  \end{center}
\end{table}

\begin{table}[!tb]
 \tiny
 \begin{center}
    \begin{tabular}{|c|c|c|c|c|c|c|c|}
      \hline
      $q$ & $\alpha$ & $\sigma_{u,t} $ & $\widetilde{VR}_{t-1}$ &
      $\widetilde{VR}_{t-2}$ & $\widetilde{VR}_{t-3}$ & $\widetilde{VR}_{t-4}$
      & Q(5) \\
      \hline
       1 &    -0.358   &     0.586   &    &    &    &    &     224.86$^{**}$ \\
       &   ( -14.982  )$^{**}$ &   ( 13.606  )$^{**}$ &    &    &    &    &       \\\cline{2-8}
       &    -0.292   &     0.596   &     0.185   &    &    &    &      71.43$^{**}$ \\
       &   ( -11.837  )$^{**}$ &   ( 14.094  )$^{**}$ &   (  8.742  )$^{**}$ &    &    &    &       \\\cline{2-8}
       &    -0.252   &     0.631   &     0.161   &     0.135   &    &    &      30.64$^{**}$ \\
       &   ( -9.979  )$^{**}$ &   ( 14.946  )$^{**}$ &   (  7.590  )$^{**}$ &   (  6.281  )$^{**}$ &    &    &       \\\cline{2-8}
       &    -0.226   &     0.640   &     0.153   &     0.120   &     0.097   &    &      11.31$^{**}$ \\
       &   ( -8.760  )$^{**}$ &   ( 15.208  )$^{**}$ &   (  7.178  )$^{**}$ &   (  5.543  )$^{**}$ &   (  4.543  )$^{**}$ &    &       \\\cline{2-8}
       &    -0.210   &     0.640   &     0.148   &     0.115   &     0.087   &     0.065   &       2.53 \\
       &   ( -7.999  )$^{**}$ &   ( 15.247  )$^{**}$ &   (  6.938  )$^{**}$ &
      (  5.325  )$^{**}$ &   (  4.065  )$^{**}$ &   (  3.063  )$^{**}$ &
      \\ 
      \hline
       2 &    -0.409   &     0.560   &    &    &    &    &     164.53$^{**}$ \\
       &   ( -18.646  )$^{**}$ &   ( 14.186  )$^{**}$ &    &    &    &    &       \\\cline{2-8}
       &    -0.348   &     0.564   &     0.148   &    &    &    &      61.97$^{**}$ \\
       &   ( -14.925  )$^{**}$ &   ( 14.437  )$^{**}$ &   (  6.983  )$^{**}$ &    &    &    &       \\\cline{2-8}
       &    -0.310   &     0.590   &     0.132   &     0.110   &    &    &      35.59$^{**}$ \\
       &   ( -12.705  )$^{**}$ &   ( 15.070  )$^{**}$ &   (  6.215  )$^{**}$ &   (  5.122  )$^{**}$ &    &    &       \\\cline{2-8}
       &    -0.269   &     0.608   &     0.124   &     0.095   &     0.124   &    &       4.63 \\
       &   ( -10.682  )$^{**}$ &   ( 15.623  )$^{**}$ &   (  5.850  )$^{**}$ &   (  4.410  )$^{**}$ &   (  5.826  )$^{**}$ &    &       \\\cline{2-8}
       &    -0.256   &     0.609   &     0.120   &     0.093   &     0.118   &     0.044   &       1.14 \\
       &   ( -9.892  )$^{**}$ &   ( 15.651  )$^{**}$ &   (  5.630  )$^{**}$ &
      (  4.313  )$^{**}$ &   (  5.527  )$^{**}$ &   (  2.065  )$^{*}$ &
      \\ 
      \hline
       3 &    -0.424   &     0.534   &    &    &    &    &      87.25$^{**}$ \\
       &   ( -20.462  )$^{**}$ &   ( 14.290  )$^{**}$ &    &    &    &    &       \\\cline{2-8}
       &    -0.382   &     0.534   &     0.099   &    &    &    &      40.04$^{**}$ \\
       &   ( -16.962  )$^{**}$ &   ( 14.350  )$^{**}$ &   (  4.660  )$^{**}$ &    &    &    &       \\\cline{2-8}
       &    -0.352   &     0.549   &     0.091   &     0.080   &    &    &      26.88$^{**}$ \\
       &   ( -14.707  )$^{**}$ &   ( 14.718  )$^{**}$ &   (  4.274  )$^{**}$ &   (  3.712  )$^{**}$ &    &    &       \\\cline{2-8}
       &    -0.313   &     0.564   &     0.087   &     0.071   &     0.104   &    &       4.89 \\
       &   ( -12.508  )$^{**}$ &   ( 15.163  )$^{**}$ &   (  4.068  )$^{**}$ &   (  3.298  )$^{**}$ &   (  4.888  )$^{**}$ &    &       \\\cline{2-8}
       &    -0.301   &     0.565   &     0.084   &     0.070   &     0.101   &     0.037   &       1.35 \\
       &   ( -11.562  )$^{**}$ &   ( 15.192  )$^{**}$ &   (  3.932  )$^{**}$ &
      (  3.262  )$^{**}$ &   (  4.721  )$^{**}$ &   (  1.717  )$^{*}$ &
      \\ 
      \hline
       4 &    -0.422   &     0.516   &    &    &    &    &      35.29$^{**}$ \\ 
       &   ( -21.169 )$^{**}$ &   ( 14.376  )$^{**}$ &    &    &    &    &       \\\cline{2-8}
       &    -0.401   &     0.515   &     0.048   &    &    &    &      22.11$^{**}$ \\
       &   ( -18.361 )$^{**}$ &   ( 14.356  )$^{**}$ &   (  2.263  )$^{*}$ &    &    &    &       \\\cline{2-8}
       &    -0.382   &     0.522   &     0.046   &     0.050   &    &    &      17.41$^{**}$ \\
       &   ( -16.262 )$^{**}$ &   ( 14.520  )$^{**}$ &   (  2.136  )$^{*}$ &   (  2.314  )$^{*}$ &    &    &       \\\cline{2-8}
       &    -0.351   &     0.534   &     0.044   &     0.046   &     0.079   &    &       5.02 \\
       &   ( -14.100  )$^{**}$ &   ( 14.835  )$^{**}$ &   (  2.054  )$^{*}$ &   (  2.157  )$^{*}$ &   (  3.672  )$^{**}$ &    &       \\\cline{2-8}
       &    -0.338   &     0.535   &     0.042   &     0.047   &     0.077   &     0.033   &       1.31 \\
       &   ( -12.967  )$^{**}$ &   ( 14.866  )$^{**}$ &   (  1.980  )$^{*}$ &
      (  2.169  )$^{*}$ &   (  3.597  )$^{**}$ &   (  1.546  ) &       \\
      \hline
       5 &    -0.413   &     0.494   &    &    &    &    &      11.39$^{*}$ \\
       &   ( -21.266  )$^{**}$ &   ( 14.127  )$^{**}$ &    &    &    &    &       \\\cline{2-8}
       &    -0.413   &     0.494   &     0.001   &    &    &    &      11.37$^{*}$ \\
       &   ( -19.335  )$^{**}$ &   ( 14.119  )$^{**}$ &   (  0.007  ) &    &    &    &       \\\cline{2-8}
       &    -0.403   &     0.497   &     0.001   &     0.024   &    &    &      10.44 \\
       &   ( -17.441  )$^{**}$ &   ( 14.166  )$^{**}$ &   ( -0.004  ) &   (  1.136  ) &    &    &       \\\cline{2-8}
       &    -0.379   &     0.506   &     0.001   &     0.024   &     0.059   &    &       3.69 \\
       &   ( -15.327  )$^{**}$ &   ( 14.377  )$^{**}$ &   ( -0.012  ) &   (  1.135  ) &   (  2.733  )$^{**}$ &    &       \\\cline{2-8}
       &    -0.367   &     0.507   &    -0.001   &     0.025   &     0.059   &     0.029   &       1.17 \\
       &   ( -14.059  )$^{**}$ &   ( 14.398  )$^{**}$ &   ( -0.056  ) &   (  1.185  ) &   (  2.730  )$^{**}$ &   (  1.340  ) &       \\
       \hline
    \end{tabular}
\caption{Estimates of model (\ref{model3}), for different values of $q$. $^{*}$ indicates $95\%$ of confidence level, $^{**}$ $99\%$ of
      confidence level.}\label{table-ar-sigmau}
  \end{center}
\end{table}

\begin{table}[!tb]
\tiny
  \begin{center}
    \begin{tabular}{|c|c|c|c|c|c|c|c|c|}
      \hline
      $q$ & $\alpha$ & $\sigma_{p,t} $ & $\sigma_{u,t}$ &  $\widetilde{VR}_{t-1}$
      & $\widetilde{VR}_{t-2}$ & $\widetilde{VR}_{t-3}$ &
      $\widetilde{VR}_{t-4}$ & Q(5) \\
      \hline
      1 &    -2.002   &    -0.167   &     0.586   &    &    &    &    &     169.25$^{**}$ \\
      &   ( -6.631  )$^{**}$ &   ( -5.462  )$^{**}$ &   ( 13.705  )$^{**}$ &    &    &    &    &       \\\cline{2-9}
      &    -1.928   &    -0.166   &     0.596   &     0.184   &    &    &    &      45.00$^{**}$ \\
      &   ( -6.506  )$^{**}$ &   ( -5.540  )$^{**}$ &   ( 14.200  )$^{**}$ &   (  8.791  )$^{**}$ &    &    &    &       \\\cline{2-9}
      &    -1.856   &    -0.163   &     0.631   &     0.161   &     0.133   &    &    &      18.54$^{**}$ \\
      &   ( -6.320  )$^{**}$ &   ( -5.483  )$^{**}$ &   ( 15.043  )$^{**}$ &   (  7.647  )$^{**}$ &   (  6.230  )$^{**}$ &    &    &       \\\cline{2-9}
      &    -1.782   &    -0.158   &     0.639   &     0.153   &     0.118   &     0.092   &    &       8.04 \\
      &   ( -6.084  )$^{**}$ &   ( -5.334  )$^{**}$ &   ( 15.292  )$^{**}$ &   (  7.247  )$^{**}$ &   (  5.521  )$^{**}$ &   (  4.363  )$^{**}$ &    &       \\\cline{2-9}
      &    -1.710   &    -0.152   &     0.639   &     0.149   &     0.114   &     0.084   &     0.058   &       3.97 \\
      &   ( -5.828  )$^{**}$ &   ( -5.134  )$^{**}$ &   ( 15.321  )$^{**}$ &
      (  7.026  )$^{**}$ &   (  5.324  )$^{**}$ &   (  3.938  )$^{**}$ &   (
      2.752  )$^{**}$ &       \\
      \hline
      2 &    -2.506   &    -0.213   &     0.560   &    &    &   &    &      95.20$^{**}$ \\
      &   ( -9.113  )$^{**}$ &   ( -7.650  )$^{**}$ &   ( 14.392  )$^{**}$ &    &    &    &    &       \\\cline{2-9}
      &    -2.371   &    -0.205   &     0.563   &     0.141   &    &    &    &      30.16$^{**}$ \\
      &   ( -8.694  )$^{**}$ &   ( -7.443  )$^{**}$ &   ( 14.631  )$^{**}$ &   (  6.758  )$^{**}$ &    &    &    &       \\\cline{2-9}
      &    -2.275   &    -0.199   &     0.588   &     0.127   &     0.103   &    &    &      20.96$^{**}$ \\
      &   ( -8.365  )$^{**}$ &   ( -7.254  )$^{**}$ &   ( 15.219  )$^{**}$ &   (  6.036  )$^{**}$ &   (  4.849  )$^{**}$ &    &    &       \\\cline{2-9}
      &    -2.148   &    -0.190   &     0.605   &     0.119   &     0.089   &     0.115   &    &       6.31 \\
      &   ( -7.928  )$^{**}$ &   ( -6.965  )$^{**}$ &   ( 15.730  )$^{**}$ &   (  5.701  )$^{**}$ &   (  4.192  )$^{**}$ &   (  5.466  )$^{**}$ &    &       \\\cline{2-9}
      &    -2.097   &    -0.186   &     0.605   &     0.116   &     0.088   &     0.111   &     0.031   &       6.59 \\
      &   ( -7.693  )$^{**}$ &   ( -6.782  )$^{**}$ &   ( 15.739  )$^{**}$ &
      (  5.542  )$^{**}$ &   (  4.130  )$^{**}$ &   (  5.255  )$^{**}$ &   (
      1.458  ) &       \\
      \hline
      3 &    -2.365   &    -0.197   &     0.534   &    &    &    &    &      43.93$^{**}$ \\
      &   ( -9.084  )$^{**}$ &   ( -7.477  )$^{**}$ &   ( 14.488  )$^{**}$ &    &    &    &    &       \\\cline{2-9}
      &    -2.283   &    -0.193   &     0.534   &     0.094   &    &    &    &      16.33$^{**}$ \\
      &   ( -8.785  )$^{**}$ &   ( -7.341  )$^{**}$ &   ( 14.542  )$^{**}$ &   (  4.444  )$^{**}$ &    &    &    &       \\\cline{2-9}
      &    -2.219   &    -0.189   &     0.548   &     0.086   &     0.074   &    &    &      14.10$^{*}$ \\
      &   ( -8.543  )$^{**}$ &   ( -7.218  )$^{**}$ &   ( 14.877  )$^{**}$ &   (  4.087  )$^{**}$ &   (  3.471  )$^{**}$ &    &    &       \\\cline{2-9}
      &    -2.125   &    -0.183   &     0.562   &     0.082   &     0.065   &     0.097   &    &       5.37 \\
      &   ( -8.199  )$^{**}$ &   ( -7.023  )$^{**}$ &   ( 15.288  )$^{**}$ &   (  3.898  )$^{**}$ &   (  3.087  )$^{**}$ &   (  4.600  )$^{**}$ &    &       \\\cline{2-9}
      &    -2.087   &    -0.180   &     0.562   &     0.080   &     0.065   &     0.095   &     0.025   &       5.33 \\
      &   ( -8.005  )$^{**}$ &   ( -6.883  )$^{**}$ &   ( 15.296  )$^{**}$ &
      (  3.807  )$^{**}$ &   (  3.071  )$^{**}$ &   (  4.487  )$^{**}$ &   (
      1.177  ) &       \\
      \hline
      4 &    -2.086   &    -0.169   &     0.516   &    &    &    &    &      15.30$^{**}$ \\
      &   ( -8.318  )$^{**}$ &   ( -6.657  )$^{**}$ &   ( 14.533  )$^{**}$ &    &    &    &    &       \\\cline{2-9}
      &    -2.056   &    -0.168   &     0.515   &     0.046   &    &    &    &       8.22 \\
      &   ( -8.191  )$^{**}$ &   ( -6.617  )$^{**}$ &   ( 14.514  )$^{**}$ &   (  2.148  )$^{*}$ &    &    &    &       \\\cline{2-9}
      &    -2.025   &    -0.166   &     0.522   &     0.043   &     0.047   &    &    &       8.49 \\
      &   ( -8.062  )$^{**}$ &   ( -6.571  )$^{**}$ &   ( 14.664  )$^{**}$ &   (  2.029  )$^{*}$ &   (  2.188  )$^{*}$ &    &    &       \\\cline{2-9}
      &    -1.972   &    -0.164   &     0.533   &     0.041   &     0.043   &     0.075   &    &       4.35 \\
      &   ( -7.860  )$^{**}$ &   ( -6.494  )$^{**}$ &   ( 14.963  )$^{**}$ &   (  1.951  )$^{*}$ &   (  2.038  )$^{*}$ &   (  3.536  )$^{**}$ &    &       \\\cline{2-9}
      &    -1.939   &    -0.162   &     0.534   &     0.040   &     0.044   &     0.074   &     0.025   &       3.73 \\
      &   ( -7.696  )$^{**}$ &   ( -6.387  )$^{**}$ &   ( 14.975  )$^{**}$ &
      (  1.897  )$^{*}$ &   (  2.055  )$^{*}$ &   (  3.480  )$^{**}$ &   (
      1.162  ) &       \\
      \hline
      5 &    -1.873   &    -0.148   &     0.494   &    &    &    &    &       4.70 \\
      &   ( -7.648  )$^{**}$ &   ( -5.980  )$^{**}$ &   ( 14.251  )$^{**}$ &    &    &    &    &       \\\cline{2-9}
      &    -1.874   &    -0.148   &     0.494   &    -0.001   &    &    &    &       4.75 \\
      &   ( -7.640  )$^{**}$ &   ( -5.979  )$^{**}$ &   ( 14.244  )$^{**}$ &   ( -0.054  ) &    &    &    &       \\\cline{2-9}
      &    -1.861   &    -0.148   &     0.497   &    -0.001   &     0.023   &    &    &       5.54 \\
      &   ( -7.582  )$^{**}$ &   ( -5.968  )$^{**}$ &   ( 14.286  )$^{**}$ &   ( -0.064  ) &   (  1.082  ) &    &    &       \\\cline{2-9}
      &    -1.830   &    -0.147   &     0.506   &    -0.002   &     0.023   &     0.057   &    &       3.68 \\
      &   ( -7.455  )$^{**}$ &   ( -5.942  )$^{**}$ &   ( 14.491  )$^{**}$ &
      ( -0.072  ) &   (  1.081  ) &   (  2.681  )$^{**}$ &    &
      \\\cline{2-9}
      &    -1.802   &    -0.145   &     0.506   &    -0.002   &     0.024   &     0.057   &     0.023   &       3.44 \\
      &   ( -7.319  )$^{**}$ &   ( -5.859  )$^{**}$ &   ( 14.499  )$^{**}$ &   ( -0.107  ) &   (  1.129  ) &   (  2.679  )$^{**}$ &   (  1.066  ) &       \\
      \hline
    \end{tabular}
\caption{Estimates of model (\ref{model4}), for different values of
      $q$. $^{*}$ indicates $95\%$ of confidence level, $^{**}$ $99\%$ of
      confidence level.}\label{table-ar-sigma-sigmau}
  \end{center}
\end{table}

\appendix
\newpage
\section{Variance Ratio asymptotic distribution}\label{appendix}
Under the null hypothesis of random walk, 
the asymptotic distribution of the statistics (\ref{vrq}) is the
following. 
Define:
\begin{eqnarray}
  \hat{\delta}_k = \frac{nq \displaystyle{\sum_{j=k+1}^{nq}} (P_j - P_{j-1} - \hat{\mu})^2(P_{j-k} - P_{j-k-1} - \hat{\mu})^2}{\Big[\displaystyle{\sum_{j=1}^{nq}} (P_j - P_{j-1} - \hat{\mu})^2\Big] ^2}\label{delta} \\
  \hat{\theta}(q) = 4 \sum_{k=1}^{q-1} \Bigg(1 - \frac{k}{q} \Bigg)^2
  \hat{\delta}_k\label{theta}.
\end{eqnarray}
Then we have:
\begin{equation}
  \sqrt{nq}(\widehat{VR}(q) - 1) \sim N(0,\hat{\theta}),\label{etero}
\end{equation}
The variance ratio test implemented here allows for
heteroskedasticity, does not require the assumption of normality and
in small samples it is more powerful than other tests, like the
Ljung-Box statistics or the Dickey-Fuller unit root test, see
\citet{LoMac89,Fau92,CecSan94}.

\end{document}